# On the combination of static analysis for software security assessment – a case study of an open-source e-government project

Anh Nguyen-Duc[1], Manh-Viet Do[2], Quan Luong-Hong[3], Kiem Nguyen-Khac[4]

[1] University of South Eastern Norway; angu@usn.no

[2] MQ ICT SOLUTIONS; dovietmanh@mqsolutions.vn

[3] MQ ICT SOLUTIONS; luong.hong.quan@mqsolutions.vn

[4] Hanoi University of Science and Technology - School of Electronics and Telecommunication; kiem.nguyenkhac@hust.edu.vn

ARTICLE INFO

ABSTRACT

Static Application Security Testing (SAST) is a popular quality assurance technique in software engineering. However, integrating SAST tools into industry-level product development and security assessment poses various technical and managerial challenges. In this work, we reported a longitudinal case study of adopting SAST as a part of a human-driven security assessment for an open-source e-government project. We described how SASTs are selected, evaluated, and combined into a novel approach for software security assessment. The approach was preliminarily evaluated using semi-structured interviews. Our result shows that (1) while some SAST tools out-perform others, it is possible to achieve better performance by combining more than one SAST tools and (2) SAST tools should be used towards a practical performance and in the combination with triangulated approaches for human-driven vulnerability assessment in real-world projects.

## 1. Introduction

Digital transformation promises to fundamentally reshape organizations and businesses by adopting cloud computing, big data, e-government [1], artificial intelligence [2], and internet-of-things [3, 4]. Before the potential benefits of adopting technologies are fulfilled, it is essential to ensure the transformation does not lead to additional harm or danger to customers and end-users. Several global, large-scale reports have shown that security and privacy continue to be major concerns to a successful digital transformation. An independent survey in 2016 reveals that almost 60 percent of participated organizations experienced a security attack and in 30 percent of them, it occurs every day [5]. Moreover, 20 percent of them are dealing with internal vulnerabilities at least quarterly. According to another report from the Identify Theft Resource Center (ITRC), data breaches in 2017 increase 45 percent than those in 2016 [6]. In another report, it is estimated that the annual cost from security issues to the global economy is more than 400 billion dollars per year [7].

Software is a common component of any digital system or service. A software vulnerability can be seen as a flaw, weakness, or even an error in the codebase that can be exploited by hackers to violate the security and privacy attributes of end-users and the software [8]. Security engineering research has investigated various approaches to identify, model, and protect software against vulnerability. One of the most common identification techniques is the static analysis of source code, which investigates the written source code to find software flaws or weak points. Static Application Security Testing (SASTs) is accepted and used in many software development companies as a gatekeeper of code quality [9,10]. There are many different SAST tools available, ranging from commercial tools to open-source projects, from using simple lexical analyses to more comprehensive and complex analysis techniques, from a stand-alone tool to a component in a development pipeline. Some popular SAST tools, such as SonarQube or IntelliJ IDEA are integral parts of continuous software development cycles. Given the wide range of SAST tools, software developers, security professionals, and auditors might face a question: how to choose a suitable SAST for their project?

One of the known problems of using SASTs is the possibly large number of misleading warnings [11, 12]. SASTs can report many false positives that attract unnecessary debugging efforts. In practice, the usefulness of SASTs depends on the project and organizational context and in general requires empirical investigation before the full adoption. The combination of SAST tools could also be an interesting approach to increase the overall effectiveness of static testing. There exist several empirical studies about the effectiveness of SASTs [9, 11-16]. Most of these studies base on test data or open-source data and do not reflect the adoption of this type of tool in a real-world context.

We conducted a case study in a large-scale government project that aims at developing a secured and open-source software system for e-government. SASTs were integrated into a so-called security gate that analyzes and assesses the vulnerability level of incoming source code. The security gate needs to ensure that all software will go through a security analysis before going further to end-users. In this project, SAST tools are experimented with and adopted to assist vulnerability assessment of open-source software.

This paper reports our experience with adopting SAST tools in the e-government project through a research-driven process. We proposed two Research Questions (RQs) to guide the development of this paper. Firstly, we would like to investigate the state-of-the-art SAST tools and the ability to combine them in detecting software vulnerabilities. Secondly, we explore an industrial experience that adopted a combination of SAST tools in supporting security assessment in an e-government project.

- RQ1: Is it possible to increase the performance of SAST tools by combining them?
  - o RQ1a. Which SAST tool has the best performance against the Juliet Test suite?
  - o RQ1b. Is the performance of SAST increased when combining different tools?
- RQ2: How SAST tools can support security assessment activities in an open-source e-government project?

The contribution of this paper is as follows:
- An overview of state-of-the-art SAST tools and their applications
- An experiment that evaluates the performance of these tools
- An in-depth case study about the adoption of SAST tools in e-government projects.

The paper is organized as follows. Section 2 presents backgrounds about software security, SAST, and security in e-government sectors. Section 3 describes our case study of the Secured Open source-software Repository for E-Government (SOREG). Two main research components of the projects are presented in this paper, an experiment with different SAST tools in Section 4 and a qualitative evaluation of a combined SAST approach for support-ing security assessment in Section 5. After that, Section 6 discusses the experience in this paper, and Section 7 concludes the paper.

## 2. Background

### 2.1. Software security and vulnerabilities

Software security is "the idea of engineering software so that it continues to function correctly under malicious attack" [17]. Factors impacting whether code is secured or not include the skills and competencies of the developers, the complexity level of the software components and its associated data, and the organizational security policies [17, 18]. We need to distinguish relevant or similar terms about software security:

- Vulnerability: can be defined as flaws or weaknesses in software design, implementation or operation management and can be exposed to break through security policies [19].
- Fault: a condition that causes the software to fail to perform its required function.
- Failure: is the deviation of actual functioning outputs from its expected outputs, or in another word, is when a fault is exploited leading to a negative consequence to the software.
- Attack: an unauthorized attempt to steal, damage, or expose systems and data via exploiting vulnerabilities.

Security experts and communities maintain different databases and taxonomies of vulnerabilities, for instance, CVE, CWE, NPD, MFSA, OWASP, and Bugzilla. The common weaknesses and enumeration (CWE) dictionary by MITRE provides common categories of software vulnerabilities. For instance, cross-site scripting (XSS) describes a type of vulnerability that occurs when form input is taken from a user and not properly validated, hence, allowing for malicious code to be injected into a web browser and subsequently displayed to end-users. SQL-Injection is another common type of vulnerability, where user input is not correctly validated and directly inserted in a database query. A path manipulation type occurs when users can view files or folders outside of those intended by the application. Buffer handling vulnerabilities allow users to exceed the buffer's bounds which can result in attacks ranging from writing instructions to gaining full system access or control. The overview of different CWE categories is given in Table 1.

Table 1: CWE categories and examples [20]

| Class Id | Weakness class | Example Weakness (CWE Entry) |
|---|---|---|
| W321 | Authentication and Access Control | CWE-285: Improper Authorization |
| W322 | Buffer Handling (C/C++ only) | CWE-120: Buffer Copy without Checking Size of Input |
| W323 | Code Quality | CWE-561: Dead Code |
| W324 | Control Flow Management | CWE-705: Incorrect Control Flow Scoping |
| W325 | Encryption and Randomness | CWE-328: Reversible One-Way Hash |
| W326 | Error Handling | CWE-755: Improper Handling of Exceptional Conditions |
| W327 | File Handling | CWE-23: Relative Path Traversal |
| W328 | Information Leaks | CWE-534: Information Exposure Through Debug Log Files |
| W329 | Injection | CWE-564: SQL Injection: |
| W3210 | Malicious Logic | CWE-506: Embedded Malicious Code |
| W3211 | Number Handling | CWE-369: Divide by Zero |
| W3212 | Pointer and Reference Handling | CWE-476: NULL Pointer Dereference |

### 2.2. Static Application Security Testing (SAST)

Security testing is an area of testing that verifies the software with the purpose of identifying any fault, attack, or failure that is different from the given security requirements. There are two major types of security testing, i.e. static testing and dynamic testing [17]. Static Application Security Testing (SAST) utilizes a static code analysis tool to analyze source code to identify potential vulnerabilities or software faults. Different from dynamic approaches, SAST examines source code without executing it, and by checking the code structure, the logic flows of statements, the usage of variables, values, functions, and procedures. Common techniques used in SAST tools include (1) syntactic analysis, such as calling insecure API functions or using insecure configuration options, and (2) semantic analysis that requires an understanding of the program semantics, i.e. data flow or control flows. This analysis starts by representing the source code by an abstract model (e.g., call graph, control-flow graph, or UML class/sequence diagram).

As SAST tools work as white box testing and do not actually run the source code, a reported vulnerability from the tool might not necessarily be an actual one. The reason for a wrong warning might because (1) the source code is secure (true negative) or (2) the source code has a vulnerability but is not reported by the tools (false negative).

Research about SAST tools is not new. The Center for Assured Software (CAS) developed a test suite with "good code" and "faulted code" across different languages to evaluate the performance of static analysis tools [16]. Tracing research work that has used this test suite, we found several existing empirical studies that assessed the effectiveness of SAST tools, in terms of accuracy, precision, and recall. Okun et al. assessed five commercial SAST tools and reported the highest recall score of 0.67 and the highest precision score of 0.45 [16]. Dıaz and Bermejo compared the performance of nine SAST tools, including commercial ones, and found an average recall value of 0.527 and an average precision value of 0.7 [21]. Charest investigated four different SAST tools in detecting a class of CWE using the Juliet test suite [22]. The best-observed performance in terms of recall was 0.46 for CWE89 with an average precision of 0.21. Baca et al. evaluated the use of a commercial SAST tool and found it is difficult to apply in an industrial setting [14]. In his case, the process of correcting false-positive findings leads to additional vulnerability in the existing secure source code. Hofer conducted few experiments and found that different SAST tools detected different kinds of weaknesses [15]. In this research, we will analyze the effectiveness of seven different SAST tools. Different from previous research, we also investigate the performance of combining these tools.

## 2.3. Vulnerability databases

It has been a worldwide effort of capturing and publishing known vulnerabilities in common software via vulnerability databases. The vulnerability databases record structured instances of vulnerabilities with their potential consequences. It helps developers and testers to be aware of and keep track of existing vulnerabilities in their developing systems [23]. According to a public report [24], there are more than 20 active vulnerability databases. Many of the databases are the results of a global effort by communities to leverage the existing large number of diverse real-world vulnerabilities. The most popular databases include:

- National Vulnerability Database (NVD)[1] - operated by the US National Institute of Standards and Technology, NVD contains known vulnerable information in the form of security checklists, vulnerability descriptions, misconfigurations, product names, and impact metrics.

- Common Vulnerabilities and Exposures (CVE)[2] – operated by the MITRE Corporation, CVE contains publicly known information-security vulnerabilities and exposures.
- Common Weakness Enumeration (CWE)[3] – sponsored by the MITRE Corporation with support from US-CERT and the National Cyber Security Division of the U.S. Department of Homeland Security, CWE is a community-developed category that provides a common language to describe software security weaknesses and classifies them based on their reported weaknesses.
- MFSA[4] or Mozilla Foundation Security Advisory, created by Mozilla, contains vulnerabilities detected by the community. Every vulnerability record is associated with a title, impact, announced, reporter, product, and fix. It also provides links to the associated files, codes and patches via the Bugzilla[5].
- OSVDB[6] - It is an independent open-source database contributed by various researchers. It currently covers about 69.885 vulnerabilities in 31.109 software.
- OWASP TOP 10[7], created by OWASP (Open Web Application Security Project) – the non-profit organization on software security, provides the annually updated list of top ten most critical security risks to web applications. The OWASP Top 10 list is based on vulnerabilities gathered from hundreds of organizations and over 10.0000 real-world applications and APIs.

## 2.4. Security concerns in e-Government

In Field et al.'s work, e-Government is defined as the combination of information technologies and organizational transformation in governmental bodies to improve their structures and operations of government [25]. The potential benefits of e-government are (1) internal and central management of governmental information, (2) increased effectiveness of governmental services, (3) better connections between citizens, businesses, and different units in public sectors [26]. Many challenges during the deployment and operation of e-Government, particularly in development countries, are reported. The challenges include inadequate digital infrastructure, a lack of skills and competencies for design, implementation, use, and management of e-government systems, and a lack of trust in the security and privacy of the systems, to name a few [27-29].

Security and privacy threats are always major concerns for operations of e-government projects [1]. If an e-government system are not well secured, security attacks may harm the system and its users at any time, leading to different types of financial, psychological and personal damages. Bélanger et al. reported common security threats to e-government systems, including Denial of Service (DoS), unauthorized network access, cross-site scripting (XSS), and penetration attacking [30]. Alshehri

---

emphasized that privacy and security must be protected to increase the user's trust while using e-government services [31]. Security measures in an e-government system can be implemented at physical, technical, or management levels [30]. From a software engineering perspective, we focused on the technical level, in which SAST plays an important role as both vulnerability detection software and security assessment tools [32].

# 3. The case study

Vietnam is a developing country with a significant investment on a country-wide digital transformation. During the last decade, several initiatives have been implemented at regional and national levels to increase the digital capacity of the government, provide e-services to some extent, develop IT infrastructure and integrate national information systems and database. Aligned with the national strategy, a government-funded project, entitled "Secured Open source-software Repository for E-Government" (SOREG) has been conducted. It is noted that the project is among several funded R&D projects towards the implementation of the whole e-government systems in a large-scale. The project was led by a domestic software company so-called MQ Solution.

We participated in the project with both passive and participant observation. The first author of the paper participated in the project as a researcher, and is responsible for the plan and conducting an experiment with SAST tools. The research design was conducted and the experiment was carried on without affecting the original project plan. The second, third, and fourth authors of the paper directly performed the experiment following a predetermined design and also participated in developing different software modules in the projects. The first part of SOREG with literature review and market research has been partly reported in our previous work [33].

## 3.1. Research design

Since the project involves both research and development activities, we will only focus on the research-relevant parts. Figure 1 describes the research process leading to the selection and evaluation of SAST tools in supporting vulnerability assessment in SOREG. In the scope of this work, we focus on the research activity; hence, the open-source repository development is not mentioned (represented as a grey box). We also exclude the research and development of Dynamic Application Security Testing (DAST) and the integration of DAST and SAST tools in this paper (the other two grey boxes in Figure 1).

This paper reports a part of the case study with two parts, an experiment that investigates SAST tools with a test suite and a qualitative evaluation of the proposed SAST solution. When the project had started, we conducted an ad hoc literature review to understand the research area of software security testing and particularly SAST and DAST tools. After that, as a feasibility analysis, we selected a set of SAST tools and conducted an experiment to evaluate the possibility of combining SAST tools. A development activity follows with the architectural design of the integrated SAST solution and prototype development. After that, we conducted a preliminary evaluation of the prototype with expert interviews.

We described which SAST tools are selected (Section 4.1), the test suite to compare them (Section 4.2), the evaluation metric (Section 4.3), and the experiment result (Section 4.4). We briefly describe the outcome of the integrated SAST solution (Section 4.5) in this paper. The analysis of semi-structured interviews is shown in Section 5.

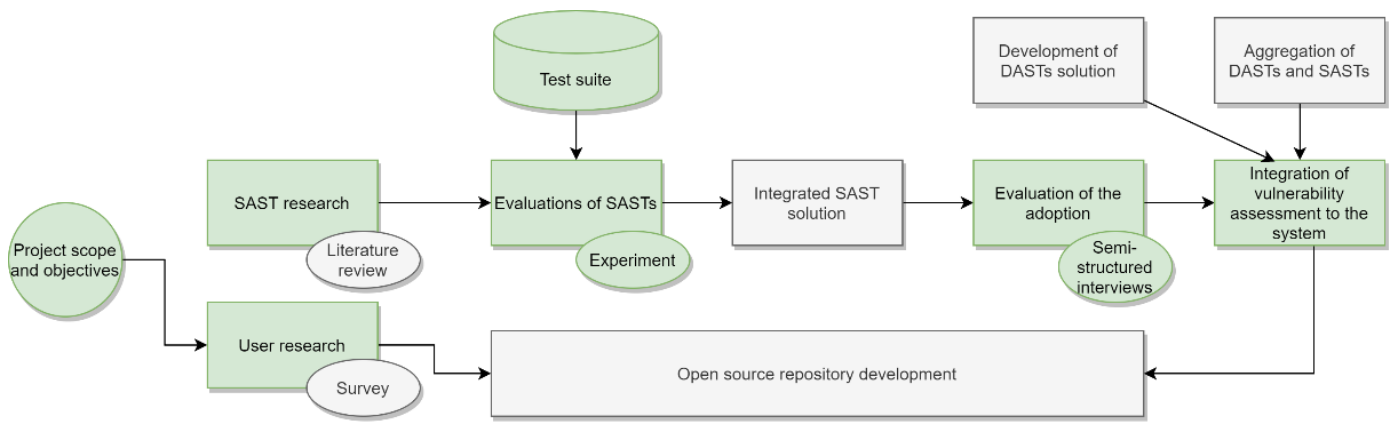

Figure 1: An overview of the research (green boxes) and development (grey boxes) process

## 3.2. SOREG – Secured Open source-software Repository for E-Government

Aligned with the e-government policy, the development and security assurance of an open-source repository is necessary. The repository will serve for ca. 2.8 million government officers. Therefore, the security aspect is of high priority. The project was funded by the Ministry of Science and Technology of Vietnam from November 2018 to February 2021. The initiated budget is 200.000 Eur. The project team includes 23 key members who participate in project planning, implementation, and closure. The main objectives of SOREG are (1) proposal and development of a prototype of a community-driven open-source software repository, (2) development and validation of a security assessment approach using SAST and DAST tools. The security assessment module focuses on software vulnerability. The expectation is that the module can detect existing vulnerabilities from dependent components, such as libraries, frameworks, plug-ins, and other software modules. Issues with X-injections, e.g. SQL injection, LDAP injection should be detected at a practical rate. Other types of vulnerabilities as described in CWE should also be covered at an acceptable level.

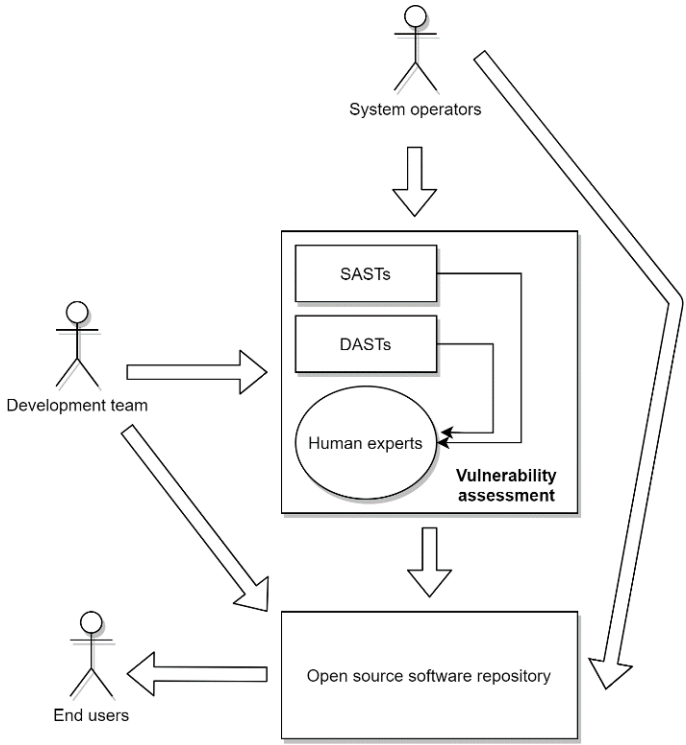

Figure 2: An overview of the open source software repository

All software submitted to the repository must go through a vulnerability assessment, as shown in Figure 2. The assessment module includes two parts (1) tools including both DAST and SAST tools, and (2) experts as moderators. The experts received reported results from tools and decide the vulnerability level of the inputted software. If the software passes the check, it will be published in the repository and available to all users. Otherwise, the software will be sent back to the submitters and not accepted for publishing.

# 4. RQ1: Is it possible to increase the performance of SAST tools by combining them?

This section presents an experiment that answers the RQ1. We explored two sub research questions:

*RQ1a. Which SAST tool has the best performance against the Juliet Test suite?*

*RQ1b. Is the performance of SAST increased when combining different tools?*

We selected seven SAST tools (Section 4.1), prepared the test suite (Section 4.2), performance metrics (Section 4.3) and reported the results for RQ1a (Section 4.4.1) and RQ1b (Section 4.4.2).

## 4.1. The selected SAST tools

The literature review on SAST and experts' opinions are the two main inputs for selecting SAST tools. We gathered seven tools that attract both research and practitioners by the time the research project was conducted (end of October 2018). The list of the tools is summarized in Table 2.

Table 2: A list of popular SAST tools by the end of 2018

| Tool name | Description |
|---|---|
| SonarQube | Scans source code for more than 20 languages for Bugs, Vulnerabilities, and Code Smells |
| Infer | Static check for C, C++, Objective-C and Java, work for iOS and Android |
| IntelliJ IDEA | integrated development environment (IDE) written in Java, support multiple languages |
| VCG | an automated code security review tool that handles C/C++, Java, C#, VB and PL/SQL |
| Huntbug | A Java bytecode static analyzer tool based on Procyon Compiler tools aimed to supersede the FindBugs |
| PMD | A cross-language static code analyzer |
| Spotbug | A static analysis for Java code, a successor of FindBugs |

### 4.1.1. SonarQube

SonarQube is one of the most common open-source static code analysis tools for measuring quality aspects of source code, including vulnerability. SonarQube implements two fundamental approaches to check for issues in source code:

- Syntax trees and API basics: Before running any rules, a code analyzer parses the given source code file and produces the syntax tree. The structure is used to clarify the problem as well as determine the strategy to use when analyzing a file.
- Semantic API: In addition to enforcing rules based on data provided by the syntax tree, SonarQube provides more information through a semantic representation of the source code. However, this model currently only works with Java source code. This semantic model provides information regarding each symbol manipulated.

Using the API, Sonar has built-in several popular and proven tools available in the open-source community. These tools, through the implementation of standardized testing source code, consider possible errors and errors, each in their own opinion. The nature of checks ranges from small styles, for example detecting unwanted gaps, to more complex spaces that are more prone to potential errors, such as variables that cannot qualify checks result in null references.

### 4.1.2. Infer

Infer, also referred to as "Facebook Infer", is a static code analysis tool developed Facebook engineers and an open-source community. Infer is a compositional program analysis, which allows feedback to be given to developers in tune with their flow of incremental development [34, 35]. Common quality checks include null pointer exceptions, resource leaks, annotation reachability, and concurrency race conditions. Infer is reportedly able to support scalable security assurance process, to run quickly and continuously on code-changes while still performing an inter-procedural analysis [36].

### 4.1.3. IntelliJ's IDEA

IntelliJ IDEA is a static code analysis feature that provides an on-the-fly code check when using IntelliJ development environment. Common vulnerabilities, inconsistencies, probable bugs, and violations can be detected during development time and later stages. The special point with IntelliJ IDEA is its source-code indexing feature that is able to produce smart quick-fixes and on-the-fly code analysis.

### 4.1.4. Visual Code Grepper (VCG)

VCG is an automated code security review tool that is applicable to many different programming languages. In addition to performing some more complex analysis, it has a portable and expandable configuration that allows users to add any bad functions (or other text). VCG offers a powerful visualization for both individual files and the codebase, showing relative proportions of code, whitespace, comments, styling comments and code smell. It is also able to identify vulnerabilities, such as buffer overflows and signed/unsigned comparisons.

### 4.1.5. Huntbug

Huntbug is a Java static analyzer tool based on Procyon Compiler tools, which aims at outperforming the famous tool FindBugs.

### 4.1.6. PMD

PMD is another popular source code analyzer that works with common programming flaws, including unused variables, empty catch blocks, and unnecessary object creation. It supports Java, JavaScript, Salesforce.com Apex and Visualforce, PLSQL, Apache Velocity, XML, XSL. Similar to SonarQube or Huntbug, PMD adopts rule-based and pattern-comparison mechanisms.

### 4.1.7. SpotBug

SpotBugs is a program that uses static analysis to look for bugs in Java code. SpotBugs checks for more than 400 bug patterns.

*4.2. Test suite*

Different open-source test suites exist for the purpose of security testing. Two popular examples are the Juliet test suite and OWASP Benchmark. We decided to use the Juliet test suite because it is not only limited to the top 10 vulnerabilities as of the OWASP benchmark dataset. In addition, the test suite covers a comprehensive list of weaknesses and supports multiple languages. One of the goals for developing the Juliet test suite was to enable open dataset for empirical research. The test suite has been popular among software and security engineering research [7, 12-14]. The latest version (ver 1.3) comprises 64.099 test cases in C/C++ and 28.881 test cases in Java[8]. In the scope of SOREG project, it is fair to focus on these two programming languages due to the dominance of them in SOREG's source code.

*4.3. Evaluation metrics*

Evaluation metrics are used extensively in empirical study about data mining, software quality predictions and software metrics. We adopted the set of evaluation metrics in our previous

work [37, 38]. These metrics are also successfully adopted in studies about security before [13, 20]:

• True Positive (TP): the reported flaw is an actual flaw (vulnerability) that needs to be fixed
• False Positive (FP): the normal non-flawed code is reported as a flaw. This warning should be ignored.
• True Negative (TN): the reported normal code is actually a non-flawed code. We do not need to do anything in this case.
• False Negative (FN): the actual flaw (embedded in the test suite) is not detected by the tool. This is a serious issue
• Recall = TP / (TP + FN)
• Precision = TP / (TP + FP)
• F1 Score = 2 (Recall x Precision)/ (Recall + Precision)

It is possible to have both TP and FP in a test file. In this case, our SAST is not sophisticated enough to discriminate for instance when the data source is hardcoded and therefore does not need to be sanitized. We adopt the "strict" metrics defined by CAS [39] as they truly reflect a real-world situation.

*4.4. Experiment Results*

### 4.4.1. RQ1.a. Which SAST tool has the best performance against the Juliet Test suite?

We report the evaluation results of the seven tools on Juliet Test Suite v1.3. as shown in Table 3. Looking at the number of outputs from each tool, IntelliJ is on top of the list with 37.694 reported issues. PMD is in second place with 37.405 reported issues and after that Sonarqube finds 28.875 reported issues. To measure the accuracy of the tools, we calculated the F1 score as shown in Table 3. The top three most accurate tools in our experiment are IntelliJ, PMD, and Sonarqube accordingly. The successors of FindBugs, i.e. Huntbugs and Spotbugs detect a small number of issues, showing their limited capacity in a software security area. Infer, the SAST promoted by Facebook finds only 428 issues from our test suite.

False Positives are also of our concern since this is one of the main barriers to adopt SAST tools in industrial projects [20]. This reflects the value of precisions. The rank of SAST is a bit different here, as Sonarqube has the best precision value (0,6), following by VCG (0,59) and IntelliJ (0,57).

We looked into details how each tool performs regarding CWE categories. Table 4 reports the F1 score of the seven tools across our 12 weakness categories.

Authentication and Authorization include vulnerabilities relating to unauthorized access to a system. IntelliJ IDEA has the best F1 score of 0.53 and followed by Sonar Qube with 0.26. Overall the ability to detect issues with SAST tools in this category is quite limited.

> Answer to RQ1a: Sonarqube has the best precision score of 0.6. IntelliJ has the best F1 score of 0.69. For a single CWE class, the best achieved F1 score is from PMD for the error handling class.

---

Table 3: Evaluation results of SAST tools against Juliet 3.1. Testsuite

| Tool | No. Detections | TP | FP | FN | Recall | Precision | F1-Score |
|------|----------------|-----|-----|-----|--------|-----------|----------|
| Sonarqube | 28875 | 9381 | 6216 | 17321 | 0.35 | 0.6 | 0.44 |
| Infer | 428 | 1564 | 1364 | 45768 | 0.03 | 0.53 | 0.06 |
| Intellij | 37694 | 52276 | 40026 | 8502 | 0.86 | 0.57 | 0.69 |
| VCG | 8143 | 8900 | 6164 | 38053 | 0.19 | 0.59 | 0.29 |
| PMD | 37405 | 12094 | 10389 | 8791 | 0.58 | 0.54 | 0.56 |
| Huntbugs | 2677 | 1873 | 2138 | 43519 | 0.04 | 0.47 | 0.07 |
| SpotBug | 624 | 347 | 313 | 45572 | 0.01 | 0.53 | 0.02 |

Table 4: Evaluation results of SAST tools across different CWE categories

| CWE Class | Sonar Quebe | Infer | Intellij IDEA | VCG | PMD | Huntbugs | Spotbugs |
|-----------|-------------|-------|---------------|-----|-----|----------|----------|
| Authentication and Access Control | 0.26 | 0.17 | 0.53 | 0.19 | 0.25 | 0.07 | 0 |
| Code quality | 0.63 | 0 | 0.83 | 0 | 0.79 | 0 | 0 |
| Control Flow Management | 0.38 | 0 | 0.69 | 0.28 | 0.65 | 0 | 0 |
| Encryption and Randomness | 0.62 | 0 | 0.6 | 0.15 | 0.68 | 0 | 0 |
| Error Handling | 0.64 | 0 | 0.73 | 0 | 0.84 | 0 | 0 |
| File Handling | 0.35 | 0 | 0.7 | 0.23 | 0.63 | 0.07 | 0 |
| Information Leaks | 0.67 | 0 | 0.62 | 0.45 | 0.63 | 0 | 0 |
| Initialization and Shutdown | 0.16 | 0 | 0.73 | 0.39 | 0.72 | 0 | 0 |
| X-Injection | 0.51 | 0 | 0.72 | 0.36 | 0.68 | 0.15 | 0.04 |
| Malicious Logic | 0.79 | 0.02 | 0.8 | 0.08 | 0.6 | 0 | 0 |
| Number Handling | 0.29 | 0.12 | 0.56 | 0.17 | 0.29 | 0 | 0 |
| Pointer and Reference Handling | 0.23 | 0 | 0.71 | 0.21 | 0.75 | 0 | 0 |

Code quality includes Issues not typically security-related but could lead to performance and maintenance issues. Intellij IDEA has the best F1 score of 0.83 and followed by PMD with 0.79. Sonar Qube has an F1 score of 0.63, which is in third place. Other tools are not able to detect any issues in this category. This can be explained by the coverage-by-design of the tools. SASTs such as SonarQube and Intellij cover not only vulnerabilities but also many other types of concerns, e.g. code smells, bugs, and hot spots.

Control Flow Management explores issues of sufficiently managing source code control flow during execution, creating conditions in which the control flow can be modified in unexpected ways. PMD has the best F1 score of 0.69 and followed by Sonarqube with an F1 score of 0.62.

Encryption and Randomness refer to a weak or improper usage of encryption algorithms. Intellij IDEA has the best F1 score of 0.53 and followed by Sonar Qube with an F1 score of 0.26. Overall the ability to detect issues with SAST tools in this category is quite limited.

Error Handling includes failure to handle errors properly that could lead to unexpected consequences. PMD has the best F1 score of 0.84 and followed by Intellij IDEA with an F1 score of 0.73.

File Handling includes checks for proper file handling when reading and writing to stored files. Intellij IDEA has the best F1 score of 0.7 and followed by PMD with an F1 score of 0.63.

Sonarqube works not so well with this category as its F1 score is only 0.35.

Information Leaks contain vulnerabilities about exposing sensitive information to an actor that is not explicitly authorized to have access to that information. SonarQube has the best F1 score of 0.67 and followed by Intellij IDEA and PMD with similar scores.

Initialization and Shutdown concern improper initializing and shutting down of resources. We see that IntelliJ IDEA has the best F1 score of 0.73 and followed by PMD with an F1 score of 0.72.

X-Injection: a malicious code injected in the network which fetched all the information from the database to the attacker. This is probably one of the most important types of vulnerability. In this category, IntelliJ IDEA is the most accurate tool with an F1 score of 0.72 followed by PMD (0.68) and Sonar Qube (0.51)

Malicious Logic concerns the Implementation of a program that performs an unauthorized or harmful action (e.g. worms, backdoors). This is also a very important type of vulnerability and probably among the most common ones. In this category, IntelliJ IDEA is the most accurate tool with an F1 score of 0.8 followed by Sonar Qube (0.79) and PMD (0.6)

Number Handling include issues with incorrect calculations, number storage, and conversion weaknesses. Intellij IDEA has the best F1 score of 0.56 and followed by PMD and Sonar Qube with the same F1 score of 0.29.

Pointer and Reference Handling issues, for example, the program obtains a value from an untrusted source, converts this value to a pointer, and dereferences the resulting pointer. PMD has the best F1 score of 0.75 and followed by PMD and Intellij IDEA with the same F1 score of 0.71.

### 4.4.2. RQ1.b. Is the performance of SAST increased when combining different tools?

We also investigated if combining various SAST tools can produce better performance. We run the combination of two, three, and four SAST tools and compare them against our performance metrics.

We wrote a script to try all possible combination and then step-wise reduction of tools. Table 5 presents our results in three categories (1) the combination that gives the best F1 score, (2) the combination that gives the best precision, and (3) the combination that produces the largest number of outputs. We reported again the result with IntelliJ IDEA as a benchmark for comparison.

The result shows that it is possible to increase some performance metrics by combining different SAST tools. However, there are none of the combinations can produce the best value for all performance metrics. Interestingly, we see that Sonarqube appears in all best combinations, even though the SAST does not perform the best among single SAST tools. From Table 5, we can say that the 5-tools combination including SonarQube, Intelliji, VCG, Infer and SpotBug would probably give the best outcome in all of our metrics. However, this is beyond our experiment.

Table 5: The effectiveness of combining various SAST tools

| Tool | TP | FP | FN | Recall | Precision | F-score | No of outputs | Type |
|------|-----|-----|-----|--------|-----------|---------|---------------|------|
| Sonarqube + Inteliji | 65927 | 49518 | 2199 | 0,97 | 0,57 | 0,72 | 43997 | Best F-Score |
| Sonarqube + SpotBug | 9722 | 6544 | 17296 | 0,36 | 0,62 | 0,45 | 28900 | Best Precision |
| Sonarqube + Inteliji + VCG + Infer | 73811 | 55692 | 2095 | 0,97 | 0,57 | 0,72 | 44101 | Largest amount of outputs |
| Inteliji IDEA | 52276 | 40026 | 8502 | 0,86 | 0,57 | 0,69 | 37694 | |

> Answer to RQ1b: Combining SAST tools does give a better performance than of a single SAST, and this depends on the performance metrics

### 4.5. The integrated SAST tools solution

The experiment produces an input for the design and development of the SAST module for the security assessment (as shown in Figure 2). We present only briefly the architectural decisions that are taken and some architectural views of the solution.

From practical aspects, there are several requirements for SAST modules from the repository development and operation team. The team emphasized five criteria while working with SAST tools:

- Result accuracy. SAST modules should produce results within a limited time and produce accurate enough according to the project stakeholders. The development team emphasizes the importance of False Positives (TPs).
- Simplicity. Key features like security check and visualizing the result should be straightforward so that users without security background can operate the tool autonomously. The user interface should be easy-to-understand and configurable to the local language.
- Vulnerability coverage. We focused on the ability to detect different categories in the CWE database.

Besides, detection of common vulnerabilities as identified by other industry standards such as OWASP Top 10 and SANS is desirable.
- Supports multiple languages. Ensure that the SAST tool supports popular programming languages such as Java and C++. It should also have a possibility to support both mobile and web development, including Python, PhP, Javascript, Objective C, and Ruby on Rails.
- Customizability. The ability to adapt the scan results to the different output format and integrate to different business logics. The tools should be highly portable and extendable to include new plug-ins or features by request.

Especially, when looking at the focus on FP, simplicity and the ability to customize, the development team had decided to select the combination of SonarQube and SpotBug as the most practical solution with SAST tools. The further development includes SpotBug plugin to a community-version SonarQube and a new SonarQube widget to customize the scanning result. The logical view and development view of the integrated solution is shown in Figure 3. The localized and customized user interface of the tool is illustrated in Figure 4.

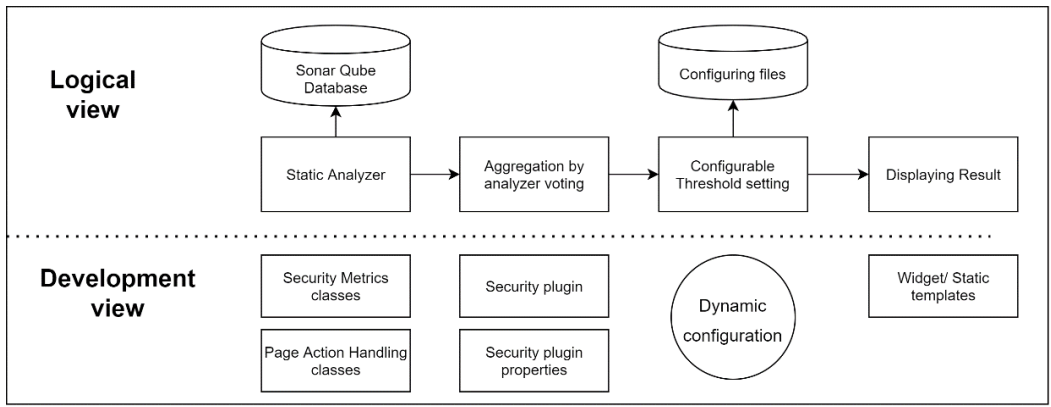

Figure 3: The logical view and development view of an integrated SAST solution

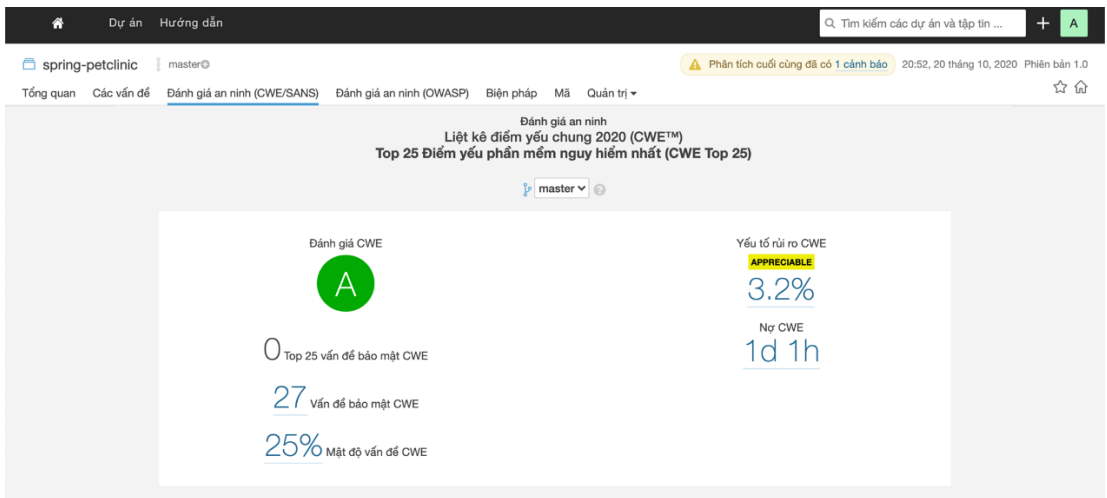

Figure 4: The UX prototype of the solution

Table 6: Profiles of interviewees

| ID | Title | Role in the project | Security expertise (1-lowest 5-highest) |
|---|---|---|---|
| P01 | Project manager | Lead the project from planning to completion. Coordinating vulnerability assessment modules with the repository | 2 |
| P02 | Security expert | Representative of an expert user of a pilot organization | 5 |
| P03 | System operator | Representative of a system operator in the pilot organization | 4 |
| P04 | Project developer | A developer of the repository | 3 |

## 5. RQ2: How SAST tools can support security assessment activities in an open-source e-government project?

After conducting the experiment (Section 4) and deciding on SonarQube and Spotbug as the foundation for static testing, the next step is to explore how the tool can be integrated into the e-government system. Semi-structured interviews were conducted with four stakeholders in the project. The profile of the interviewee is given in Table 6. The same interview guideline was used, including three main sections (1) Questions regarding the usability of the tool, (2) Questions regarding the usefulness of the tools for vulnerability assessment, and (3) Questions regarding the performance of the tools. The interviews ranged from 20 to 30 mins in total. We did note-taking during the interviews and summarized them into three main themes. It is noted that all interviews were done in Vietnamese, so the quotes are translated into English.

Perceptions about the usability of the SAST tools: there has been a consensus among interviewees about the usability of the tool. The user interface has been continuously improved and the final version has been tested with different stakeholders in the project. Before the tool's effectiveness can be evaluated, it is practically important that it can be accessed and operable by targeted users. Some feedback about the interfaces are:

*"The user interface is lean and intuitive!"* (P04)

*"I think the web interface looks good. It is easy to use and for me all the key features are visible. I do not need any manual documents to work with this tool"* (P02)

Perceptions about the performance of the SAST tools: the performance of the tools originally refers to the technical capacity of the tools against real-world applications. Our interviewee highlighted the importance of showing the possibility of capture vulnerabilities across different categories of weakness:

*"SonarQube has a wide range of test coverage. The adopted version inherited this from the community version and covers different types of vulnerabilities. It is important to be aware of different possible threats to our repository"* (P03)

We also see that performance is interpreted as a practical performance, that is the tools should be an indicator of security level and can be integrated into other ways of security assessment:

*"The experiments show that the effectiveness of top 3 SAST tools does not differ much from each other. Then we care about how difficult it is and how much time it takes to develop and integrate the selected SAST tools to our repository. The proposed architecture looks great and integral into the overall system."* (P04)

*"Testing SAST tools is an important step that giving us confidence in adopting the right tool in the next step. Performance and coverage are important insights for expert teams to decide the security level of software apps"* (P01)

Perceptions about the usefulness of the SAST tools: in the nutshell, it seems that the automated tools will not be fully automatically operated in this project. The tools are perceived as useful as a complementary means to assess security. It should be combined with another type of security testing, i.e. DAST and other types of security assessment to give a triangulated result.

*"SAST or even the combination of SAST tools and DASTs and other automated tools are not sufficient to ensure a safe software repository. I think the tools play important roles as inputs for expert teams who operate and manage security aspects of the repository."* (P03)

*"I think the tool has a good potential! I am looking forward to seeing how DASTs and SAST tools can be combined in this project"* (P02)

> Answer to RQ2: SAST tools should be used towards a practical performance and in the combination with triangulated approaches for human-driven vulnerability assessment in real-world projects.

# 6. Discussion

The experiment conducted in this research strengthens the findings from previous empirical studies on SAST tools. We updated the research of SAST tools with the state of the art tool list in 2018. By this time, we still see that using one SAST tool is not enough to cover the whole range of security weaknesses at the implementation phase. This aligns with the observation by Oyetoyan et al. [20]. However, the current SAST tools are rich in their features, e.g. ability to support multiple languages, various visualization options, and customizability. Previous studies reported the best precisions values of SAST tools around 0.45 – 0.7 [13, 15, 16]. Our study also reported the precision values of our best tools in this range. We also observed that SAST tools work relatively better in some CWE categories, such as Code quality, Encryption and Randomness, Error Handling, Information Leaks, and Malicious Logic.

In addition to existing studies, we revealed the possibility of combining different SAST tools to achieve better performance. In particular, we have used SonarQube and the base platform and combine the rulesets from other tools. As static analysis uses basically whitebox testing to explore source code, this result shows the potential to improve existing tools, and probably towards a universal security static security ruleset.

Previous studies reported many challenges in adopting SAST tools during different stages of software project life cycles [11-13, 20]. In this study, we focus on the deployment stage where software from other parties is tested before publishing. This quality check gate is common in all software repository models, such as Apple Store or Google Play. The objective of SAST here is different; we aim at supporting security assessment, not guiding software developers to improve their source code. Within the scope of SOREG, we see that the adopted approach is practically useful and contribute to the overall project scope.

Our research also has some limitations. Firstly, we only include open-source SAST tools and only conduct security testing for open-source software. However, as we see from existing research, it might not be too much different in terms of tool performance with commercial SAST tools. Secondly, our research aims at developing a prototype and evaluating the proposed solution, therefore, we did not focus on architectural details and implementation. It could be that the perceived usefulness can be improved when we have a full-scale development of the combined SAST solution. Thirdly, we preliminary tested our SAST tools with a simple software application. The claimed results are mainly based on our experiments with the Juliet test suite. A case study with a large-scale industrial application might provide more insight that is complementary to our findings. Last but not least, the study is conducted in the context of a Vietnamese government project, which would have certain unique characteristics regarding organizational and managerial aspects. However, the research design was conducted separately in Norway and the observation process has been conducted with scientific and professional attitudes. The experiment data is available published at https://docs.google.com/spreadsheets/d/18FXJKWsO6m9Ac9_-92aGmDqMd6QdFLK_rIZ9P7fF_wQ/edit#gid=2107814246. We tried to report as detail as possible the experiment process so that one can replicate our work in the same condition.

# 7. Conclusions

We have conducted a case study on a 2-year project that develop and evaluate a secured open-source software repository for the Vietnamese government. This paper reports a part of the paper about evaluating and combining SAST tools for security assessment. Among evaluated SAST tools, we have found that Sonarqube and Intellij have the best performance. The combination does give a better performance than a single SAST, depending on the performance metrics. Practically, these SAST tools should be used towards a practical performance and in the combination with triangulated approaches for human-driven vulnerability assessment in real-world projects. In the future work, we will report the next step of the project, with a similar investigation on DAST and the effectiveness of combining SAST tools and DASTs in supporting software security assessment.

## Acknowledgement

This work was co-funded under the Vietnam national project entitled "SOREG – Secured Open source-software Repository for E-Government". The project is led by MQ Solution[9].